\documentclass[%
 preprint,
nofootinbib,
 amsmath,amssymb,
 prc,
 superscriptaddress
]{revtex4-1}

\usepackage{graphicx}
\usepackage{dcolumn}
\usepackage{bm}
\usepackage{caption}
\usepackage{subcaption}
\usepackage{color}

\begin{document}


\title{Charged Particle Emissions in High-Frequency Alternative Electric Fields}


\author{Dong Bai}
\email{dbai@itp.ac.cn}
\affiliation{School of Physics, Nanjing University, Nanjing, 210093, China}%

\author{Daming Deng}
\email{dengdaming@smail.nju.edu.cn}
\affiliation{School of Physics, Nanjing University, Nanjing, 210093, China}%

\author{Zhongzhou Ren}
\email{Corresponding Author: zren@tongji.edu.cn}
\affiliation{School of Physics Science and Engineering, Tongji University, Shanghai 200092, China}%


\date{\today}

\begin{abstract}
Proton emission, $\alpha$ decay, and cluster radioactivity play an important role in nuclear physics. We show that high-frequency alternative electric fields could deform Coulomb barriers that trap the charged particle, and raise the possibility of speeding up charged particle emissions. They could also cause anisotropic effects in charged particle emissions, and introduce additional terms in the Geiger-Nuttall laws. Our study may further suggest that, for proton emitters like $^{166}$Ir, when the electric field is strong, the dominant decay mode could be changed from $\alpha$ decay to proton emission. As high-frequency alternative electric fields correspond to high-frequency laser fields in the dipole approximation, our study could be viewed as a benchmark for future theoretical studies of charged particle emissions in realistic laser fields.     
\end{abstract}

\maketitle


Recent years witness great progress in studying proton emission, $\alpha$ decay, and cluster radioactivity \cite{Delion:2010,Peonaru:2010,Pfutzner:2011ju,Delion:2018rrl}. Historically, modern theoretical nuclear physics originates from the explanation of $\alpha$ decay by Gamow, Gurney and Condon in 1928 \cite{Gamow:1928,Gurney:1928}. Interests in alpha decay persist after that, and lots of interesting results have been obtained \cite{Buck:1990zz,Buck:1992zz,Buck:1993sku,Buck:1996zza,Varga:1992zz,Delion:1992zz,Delion:2006zq,Delion:2007zz,Peltonen:2008zz,Delion:2009jw,Delion:2015wro,Delion:2015mnw,Xu:2006fq,Ni:2009vzd,Ni:2009zza,Ni:2009zzb,Ni:2009zz,Ni:2010zza,Ni:2010zzc,Ropke:2014wsa,Xu:2015pvv,Denisov:2005ax,Denisov:2009ng,Denisov:2009zza,Denisov:2015wka,Ismail:2013bma,Ismail:2015ufa,Ismail:2016zei,Mohr:2006qm,Mohr:2016ofn}.  Later discoveries of proton emission in the 1960s \cite{Jackson:1970wid,Cerny:1970zvr} and cluster radioactivity in the 1980s \cite{Sandulescu:1980,Rose:1984zz} also make important contributions to deepening our understanding of nuclei lying near the border of nuclear stability. Recently, it is pointed out by Ref.~\cite{Delion:2006nh,Ni:2008zza,Qi:2009id,Poenaru:2011zz,Qi:2012pf,Ren:2012zza,Qian:2016teg} that, $\alpha$ decay, proton emission, and cluster radioactivity could be described systematically using unified decay rules. For a pedagogic introduction to charged particle emissions, we would like to recommend Ref.~\cite{Delion:2010}. In the last few years, several works \cite{Cortes:2011,Cortes:2012tu,Misicu:2013,Kopytin:2014,Misicu:2016,Delion:2017ozx,Kis:2018llv} are devoted to $\alpha$ decay in strong electromagnetic fields, partially inspired by the upcoming powerful laser facilities in the near future \cite{ELI,ELINP}. These studies provide some preliminary hints that strong laser fields could speed up $\alpha$ decays. This is not only interesting from the pure academic viewpoint, but also might be helpful for decontaminating $\alpha$-radioactive nuclear wastes. In this note, we study charged particle emissions in high-frequency alternative electric fields, treating $\alpha$ decay, proton emission, cluster radioactivity in a unified approach inspired by Ref.~\cite{Delion:2006nh,Ni:2008zza,Qi:2009id,Poenaru:2011zz,Qi:2012pf,Qian:2016teg,Ren:2012zza}. By high-frequency alternative electric fields, we refer to alternative electric fields with frequencies (photon energies $\hbar\omega$) much higher than the $Q$ values of charged particle emissions, which means $\hbar\omega\gg Q_\alpha\sim10\text{ MeV},Q_p\sim\text{1 MeV}, Q_c\sim\text{50 MeV}$ for $\alpha$ decay, proton emission, and cluster radioactivity, respectively. High-frequency alternative electric fields correspond approximately to high-frequency laser fields in the dipole approximation. Therefore, our study could be viewed as a benchmark for future theoretical studies of charged particle emissions in realistic laser fields.

We start with the time-dependent Schr\"{o}dinger equation
\begin{align}
\!\!i\hbar\frac{\partial\Psi(\mathbf{r},t)}{\partial t}=\left[\frac{1}{2\mu}\left(\mathbf{P}-{Q_\text{eff}}\,\mathbf{A}(t)\right)^2\!\!+\!V(\mathbf{r})\right]\Psi(\mathbf{r},t),
\label{SchrodingerEquation}
\end{align} 
which describes the relative motion of the cluster and daughter nuclei in electromagnetic fields, with $\mathbf{A}(t)=\mathbf{A}_0\sin\omega t$ giving rise to the alternative electric field \cite{Joachain:2011}. $\mu=M_cM_d/(M_c+M_d)$ is the reduced mass of the two-body system, $Q_\text{eff}=eZ_\text{eff}=e(Z_c A_d-Z_d A_c)/(A_c+A_d)$ is the effective charge \cite{Misicu:2013}, and $V(\mathbf{r})$ is the original Coulomb potential between the cluster and daughter nuclei.  Here, for simplicity, we adopt the natural unit $c=1$. With the help of the Hennenberger transformation \cite{Henneberger:1968}
\begin{align}
\Omega_h(t)=\exp\left[\frac{i}{\hbar}\int_{-\infty}^t\!\!\left(-\frac{Q_\text{eff}}{\mu}\mathbf{A}\cdot\mathbf{P}+\frac{Q_\text{eff}^2}{2\mu}\mathbf{A}^2\right)\mathrm{d}\tau\right],
\end{align}
one obtains a Schr\"{o}dinger-like equation for the new wave function $\Phi=\Omega_h(t)\Psi$
\begin{align}
&i\hbar\frac{\partial\Phi(\mathbf{r},t)}{\partial t}=\left[\frac{1}{2\mu}\mathbf{P}^2+V(\mathbf{r}-\mathbf{S}(t))\right]\Phi(\mathbf{r},t),\nonumber\\
&\mathbf{S}(t)=\frac{Q_\text{eff}}{\mu}\int^t_{-\infty}\!\!\mathbf{A}(\tau)\mathrm{d}\tau.
\label{HEQ}
\end{align}
We shall call the time-dependent potential $V(\mathbf{r}-\mathbf{S}(t))$ as the Hennenberger potential for convenience, and follow the convention of laser-atom physics and call $\mathbf{S}(t)$ as the quiver displacement for the charged particle moving in alternative electric fields \cite{Joachain:2011}. 

We expand the Hennenberger potential $V(\mathbf{r}-\mathbf{S}(t))$ in terms of Fourier series. It is well-established in theoretical laser-atom physics that it is the static component that dominates over the rest Fourier components in the case of high-frequency alternative electric fields \cite{Joachain:2011}. Explicitly, the static component is given by
\begin{align}
V_0(\mathbf{r})=\frac{1}{T}\int_0^TV(\mathbf{r}-\mathbf{S}(t))\mathrm{d}t.
\end{align}
The nuclear interactions between cluster and daughter nucleus vary from nucleus to nucleus for $\alpha$ decay, proton emission, and cluster radioactivity. To give a unified treatment of all three kinds of charged particle emissions, we take into account only the Coulomb and centrifugal potentials which by themselves often could already give reliable order-of-magnitude estimates on barrier penetrabilities. We consider the possibility that the daughter or cluster nucleus may be axially deformed. The Coulomb potential is then given by \cite{Wong:1973zz}
\begin{align}
V(\mathbf{r},\Theta,\theta_\lambda)&=\frac{Z_cZ_de^2}{r}+\frac{3}{5}\frac{Z_cZ_de^2}{r^3}R_{\lambda0}^2\beta_{2\lambda}Y_{20}(\theta_\lambda)\nonumber\\
&=\frac{Z_cZ_de^2}{r}+\frac{3}{10}\sqrt{\frac{5}{4\pi}}\frac{Z_cZ_de^2}{r^5}R_{\lambda0}^2\beta_{2\lambda}\left(3{z_\lambda^2}-r^2\right),
\label{CoulombPotential}
\end{align}
with $\lambda=c$ for the cluster nucleus being axially deformed and $\lambda=d$ for the daughter nucleus being axially deformed. $\theta_d=z_d/r$ and $\theta_c=z_c/r$ $(z_\lambda=\mathbf{r}\cdot\mathbf{e}_{z_\lambda})$ measure the angles between $\mathbf{r}$ and the symmetric axis of the daughter and cluster nuclei. $R_{d0}=1.2A_d^{1/3}$ and $R_{c0}=1.2A_c^{1/3}$ are their radii. $\beta_{2d}$ and $\beta_{2c}$ are the corresponding deformation parameters. The spherical Coulomb potential could be restored by setting $\beta_\lambda=0$.

With the Coulomb potential in Eq.~\eqref{CoulombPotential}, the static component $V_0(\mathbf{r})$ is then given by
\begin{align}
&V_0(r,\Theta,\theta_\lambda,S_0)=\frac{Z_cZ_de^2}{r}\Xi_d(r,\Theta,\theta_\lambda,S_0),\nonumber\\
&\Xi_d(r,\Theta,\theta_\lambda,S_0)=\xi_{0}(r,\Theta,S_0)+\xi_{2\lambda}^{(1)}(r,\Theta,\theta_\lambda,S_0)+\xi_{2\lambda}^{(2)}(r,\Theta,\theta_\lambda,S_0),\nonumber\\
&\xi_{0}(r,\Theta,S_0)=\frac{1}{2\pi}\int_0^{2\pi}\!\!\!\!\frac{\mathrm{d}x}{\left[1-2\frac{S(x)}{r}\cos\Theta+\frac{S(x)^2}{r^2}\right]^{1/2}},\nonumber\\
&\xi_{2\lambda}^{(1)}(r,\Theta,\theta_\lambda,S_0)=\frac{9}{20\pi}\sqrt{\frac{5}{4\pi}}\frac{R_{\lambda0}^2\beta_{2\lambda}}{r^2}\int_0^{2\pi}\!\mathrm{d}x\,\frac{\left[\cos\theta_\lambda-\frac{S(x)}{r}\cos(\theta_\lambda+\Theta)\right]^2}{\left[1-2\frac{S(x)}{r}\cos\Theta+\frac{S(x)^2}{r^2}\right]^{5/2}},\nonumber\\
&\xi_{2\lambda}^{(2)}(r,\Theta,\theta_\lambda,S_0)=-\frac{3}{20\pi}\sqrt{\frac{5}{4\pi}}\frac{R_{\lambda0}^2\beta_{2\lambda}}{r^2}\int_0^{2\pi}\!\!\!\!\frac{\mathrm{d}x}{\left[1-2\frac{S(x)}{r}\cos\Theta+\frac{S(x)^2}{r^2}\right]^{3/2}},
\end{align}
with $\Theta$ being the angle between $\mathbf{r}$ and $\mathbf{S}(x)$. As before, $\lambda=c$ for axially deformed cluster nucleus, and $\lambda=d$ for axially deformed daughter nucleus. $\mathbf{S}(x)=\mathbf{S}_0\sin\omega t$, and $x=\omega t$ being the integration variable. $S_0$ is the quiver amplitude given by
\begin{align}
S_0=Z_\text{eff}\frac{\sqrt{4\pi\hbar\alpha I}}{\mu\omega^2},\label{QuiverAmplitude}
\end{align}
with $\alpha$ being the fine structure constant and $I$ being the field intensity.

According to the WKB approximation, the differential penetrability is given by
\begin{align}
&P(\Theta,\theta_\lambda)=\exp\left(-2\int_{R_t(\theta_\lambda)}^{R(\Theta)}\sqrt{\frac{2\mu}{\hbar^2}\left[U(r)-Q\right]}\mathrm{d}r\right),\\
&U(r)=\frac{\hbar^2}{2\mu r^2}L(L+1)+V_0(r,\Theta,\theta_\lambda,S_0),\label{Ur}
\end{align}
where the first term in Eq.~\eqref{Ur} is the centrifugal potential. $R_t$ is the geometric touching radius given by $R_t(\theta_c)=R_{c0}+R_{d0}+R_{c0}\beta_{2c}Y_{20}(\theta_c)$ for the cluster nucleus being axially deformed and $R_t(\theta_d)=R_{c0}+R_{d0}+R_{d0}\beta_{2d}Y_{20}(\theta_d)$ for the daughter nucleus being axially deformed. $R(\Theta)$ gives the external turning point. The total penetrability could be obtained by using the semi-classical formula \cite{Delion:2010}
\begin{align}
P=\frac{1}{4}\int_0^\pi\mathrm{d}\theta_\lambda\sin\theta_\lambda\int_0^\pi\mathrm{d}\Theta\sin\Theta\, P(\Theta,\theta_\lambda).
\end{align}
An elegant discussion on the semi-classical approach to the $\alpha$-decay rate could also be found in Ref.~\cite{Delion:2015mnw}.

\begin{figure}
\centering
\begin{subfigure}[b]{10cm}
\centering
\includegraphics[width=10cm]{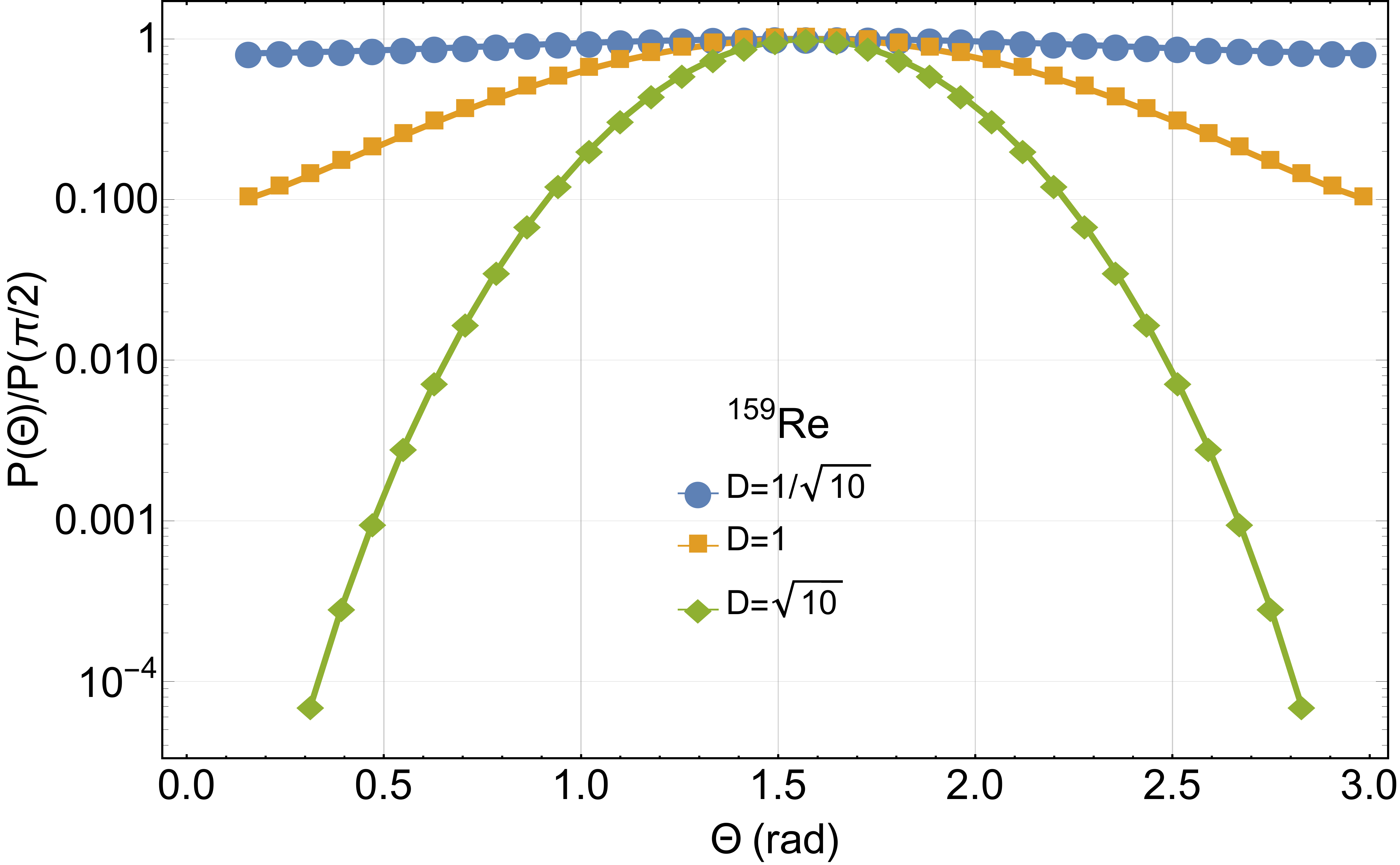}
\caption*{$\quad\quad\quad$(a)}
\end{subfigure}

\begin{subfigure}[b]{10cm}
\centering
\includegraphics[width=10cm]{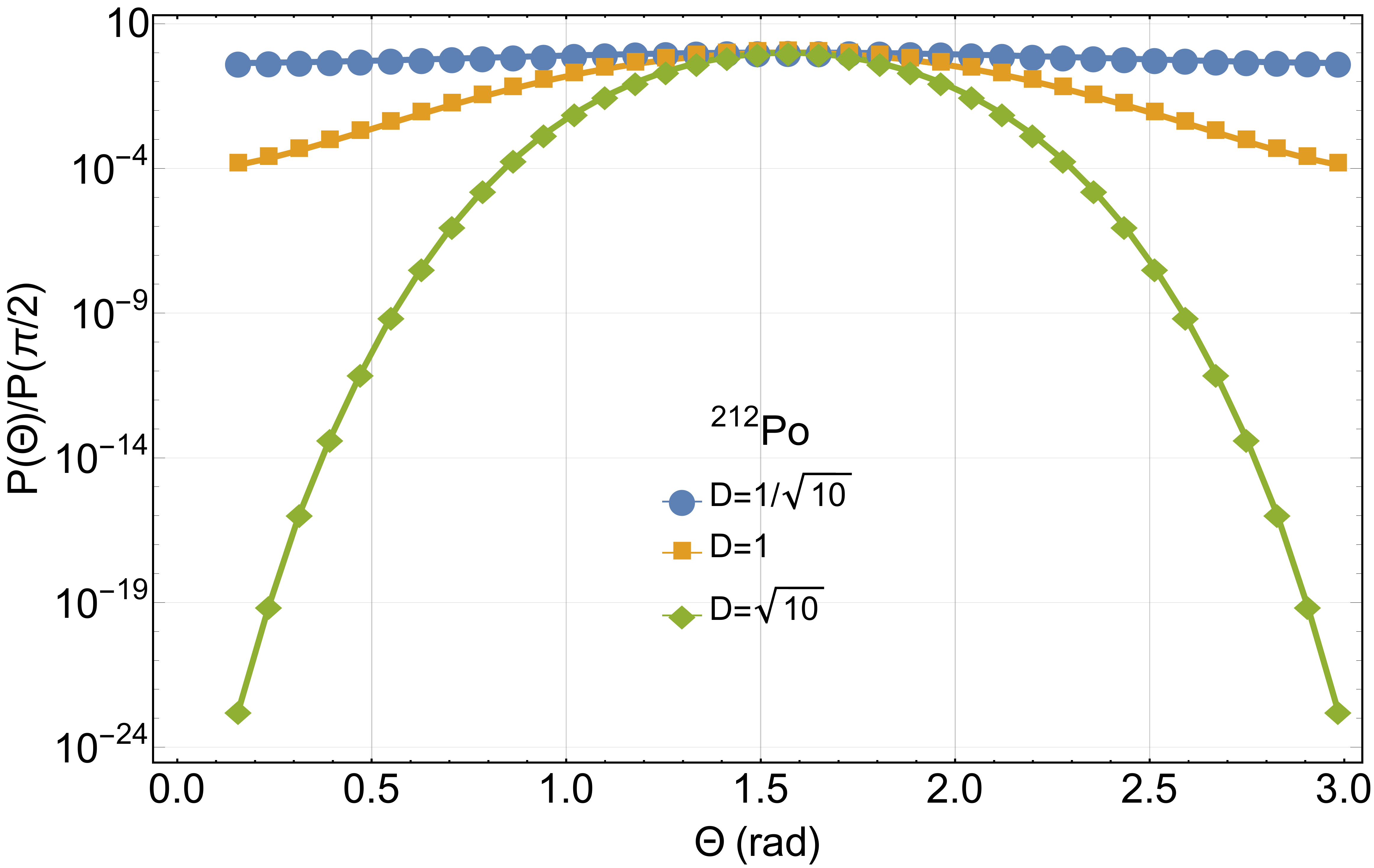}
\caption*{$\quad\quad\quad$(b)}
\end{subfigure}

\begin{subfigure}[b]{10cm}
\centering
\includegraphics[width=10cm]{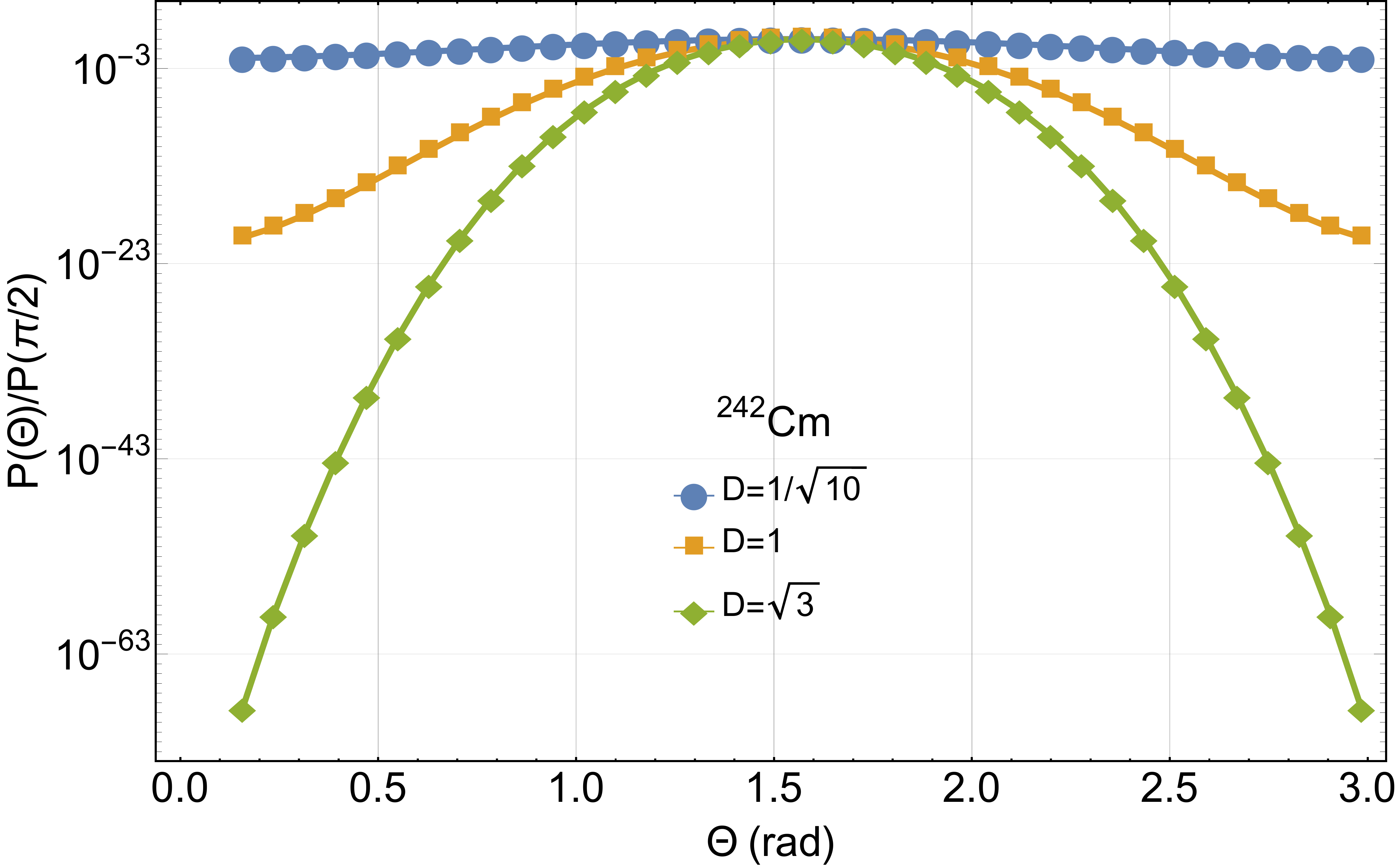}
\caption*{$\quad\quad\quad$(c)}
\end{subfigure}

\caption{Differential barrier penetrability normalized to its maximal value at $\Theta=\pi/2$ with different values of $D$ for (a) the proton emitter $^{159}$Re with the orbital angular momentum $L=5$, (b) the $\alpha$ emitter $^{212}$Po, and (c) the $^{34}$Si emitter $^{242}$Cm.}
\label{PenetrabilityIntensity}
\end{figure}

In the rest part of this note, we apply the above formalism to study the influences of high-frequency alternative electric fields on proton emission, $\alpha$ decay, and cluster radioactivity. For simplicity, we assume that the reduced widths and the $Q$ values of charged particle emissions remain approximately unchanged in high-frequency electric fields. First, we study the anisotropic effects induced by the electric fields. We pick $^{159}$Re with the orbital angular momentum $L=5$ as a representative proton emitter, ${}^{212}$Po as a representative $\alpha$ emitter, and ${}^{242}$Cm as a representative ${}^{34}$Si emitter, and treat all the daughter and cluster nuclei as approximately spherical for simplicity. Noticeably, the $\alpha$ emitter $^{212}$Po has also been studied in Ref.~\cite{Delion:2017ozx}. Following Ref.~\cite{Delion:2017ozx}, we introduce the adimensional parameter $D=S_0/R_{d0}$. Numerical results are presented in Fig.~\ref{PenetrabilityIntensity}, where the differential penetrabilities for the proton emitter $^{159}$Re, $\alpha$ emitter $^{212}$Po, and ${}^{34}$Si emitter $^{242}$Cm normalized to their maximum values are plotted. From Fig.~\ref{PenetrabilityIntensity} one could see that, when the $D$ values are tuned to be the same, the anisotropic effect is the weakest for proton emission and the strongest for heavy cluster emission. For example, given $D=1$, the differential penetrability $P(\Theta)$ varies by around one order of magnitude from $\Theta=0$ to $\Theta=\pi/2$ for the proton emitter ${}^{159}$Re, around four orders of magnitude for the $\alpha$ emitter ${}^{212}$Po, and more than twenty orders of magnitude for the ${}^{34}$Si emitter ${}^{242}$Cm.

\begin{figure}
\centering
\begin{subfigure}[b]{\textwidth}
\centering
\minipage{0.5\textwidth}
  \includegraphics[width=\linewidth]{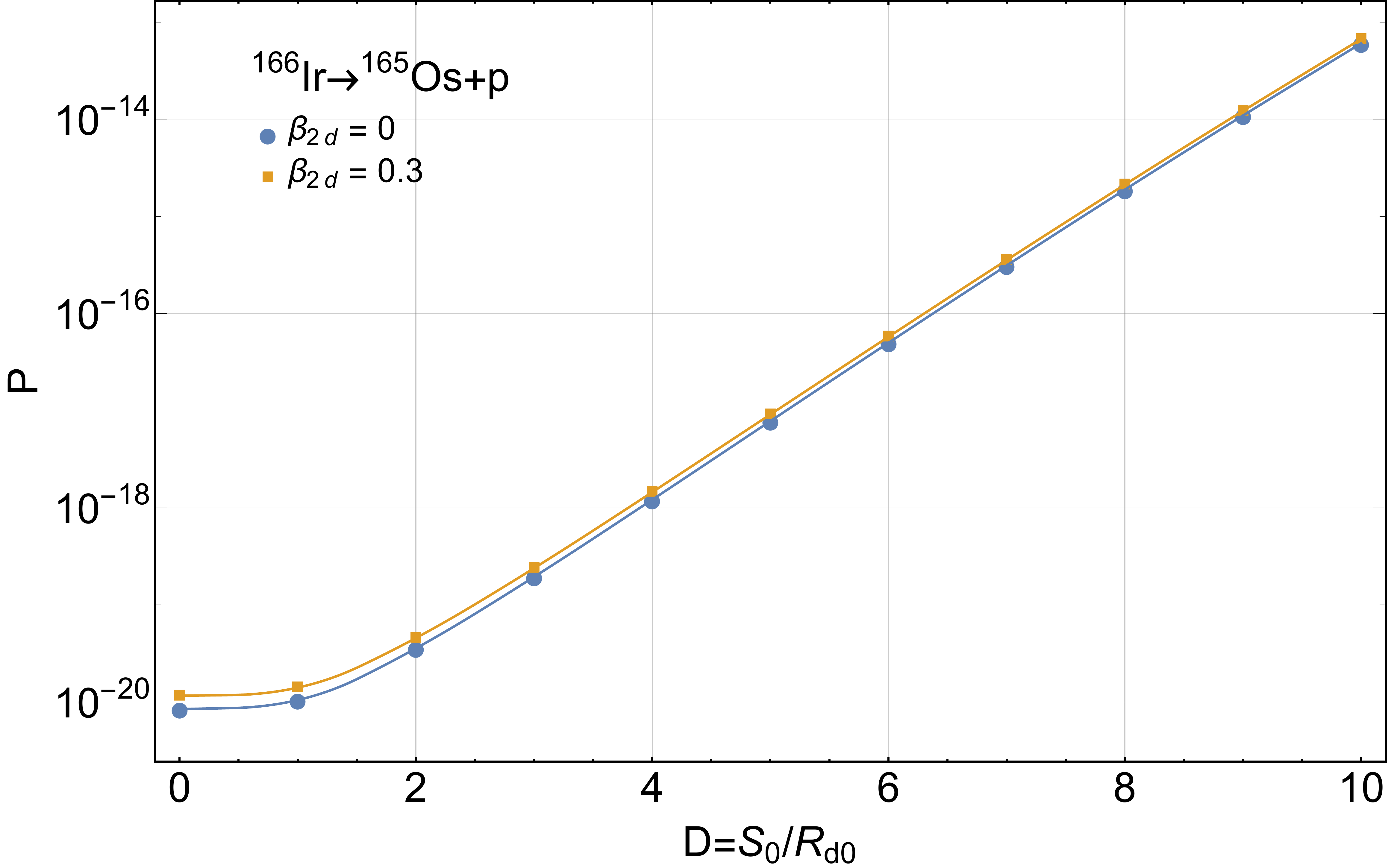}
  \caption*{$\quad\ \ \ $(a)}
\endminipage\hfill
\minipage{0.5\textwidth}
  \includegraphics[width=\linewidth]{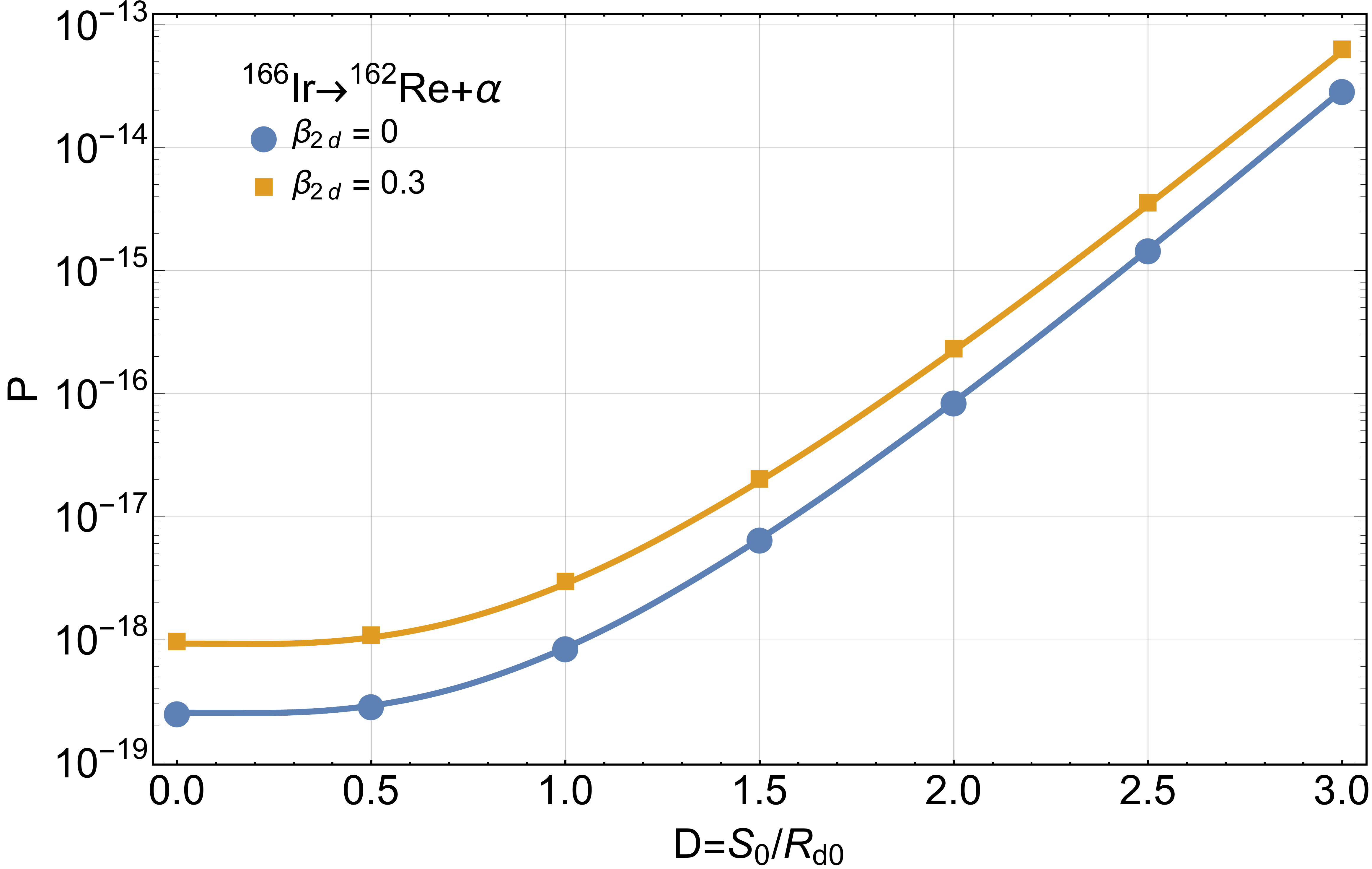}
  \caption*{$\quad\ \ \ $(b)}
\endminipage
\end{subfigure}


\begin{subfigure}[b]{\textwidth}
\centering
\minipage{0.5\textwidth}
  \includegraphics[width=\linewidth]{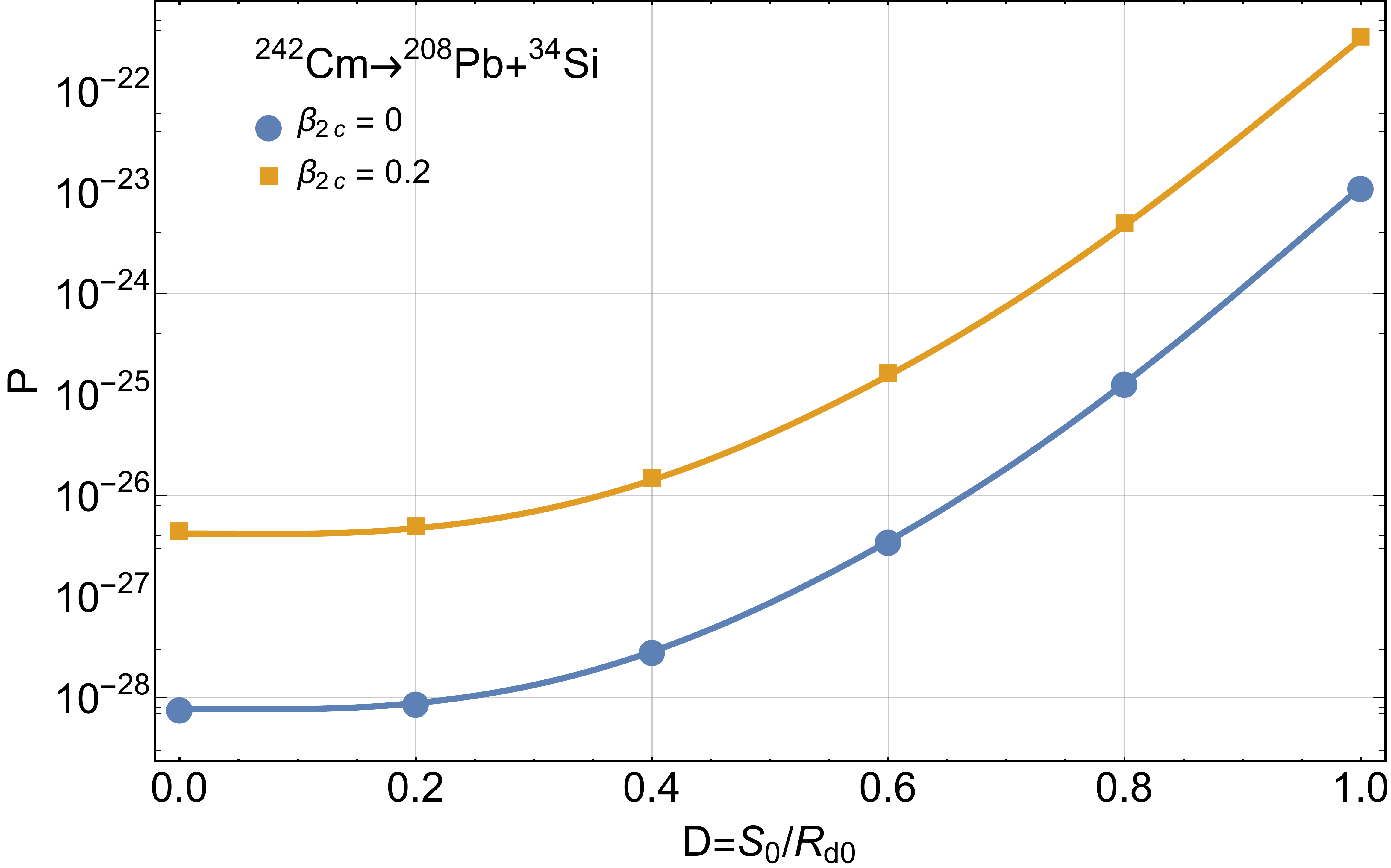}
  \caption*{$\quad\ \ \ $(c)}
\endminipage\hfill
\minipage{0.5\textwidth}
  \includegraphics[width=\linewidth]{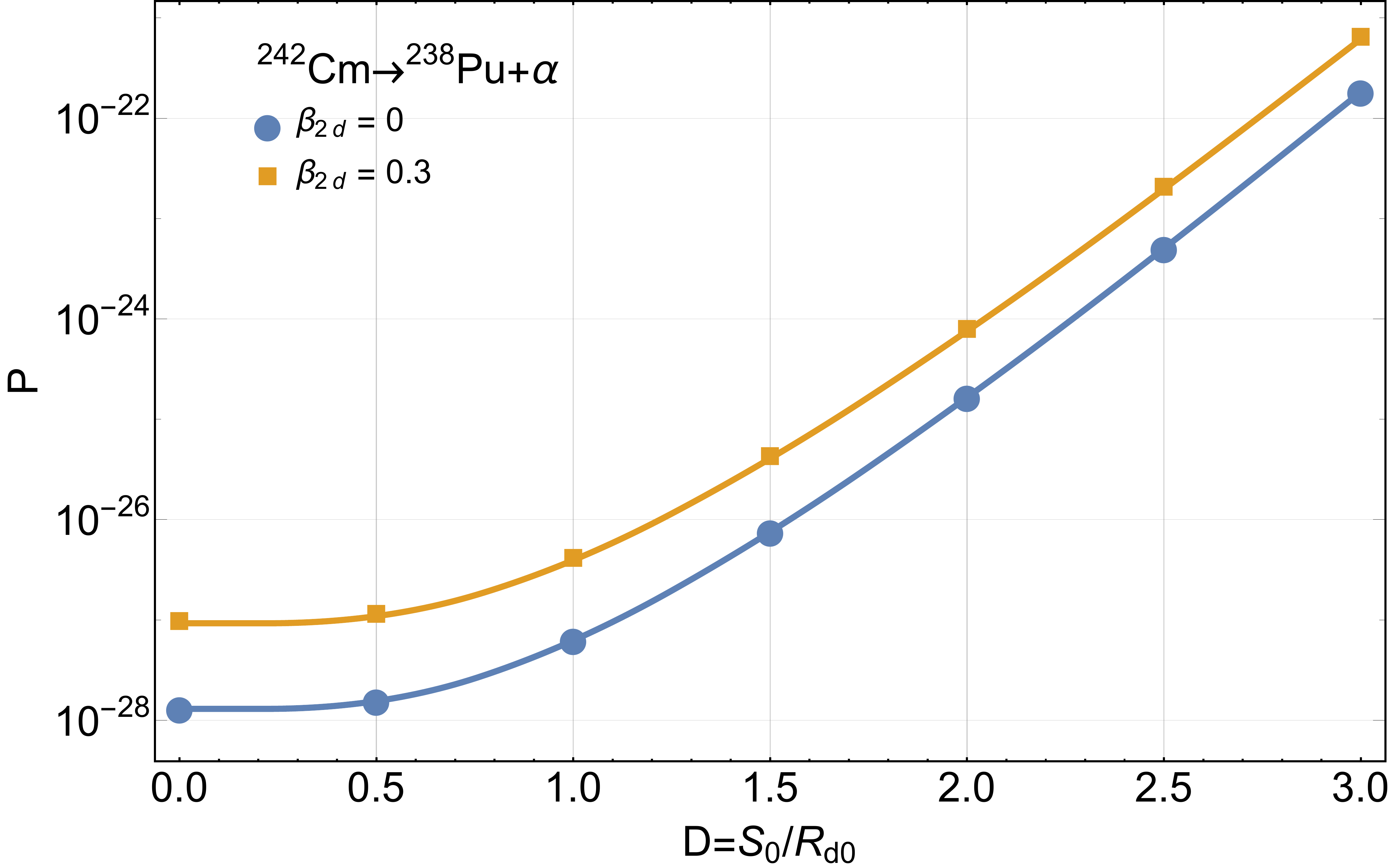}
  \caption*{$\quad\ \ \ $(d)}
  \endminipage
\end{subfigure}

\caption{Total penetrability versus $D=S_0/R_{d0}$ with different assumptions on their deformations for (a) the proton-emission channel of $^{166}$Ir with the orbital angular momentum given by $L=2$, (b) the $\alpha$-decay channel of $^{166}$Ir, (c) the $^{34}$Si-radioactivity channel of $^{242}$Cm, and (d) the $\alpha$-decay channel of $^{242}$Cm.}
\label{DCrit}
\end{figure}

We then try to determine the critical value $D_\text{crit}$ beyond which the total penetrability gets a strong increase, and study the influence of nuclear deformations. For these purposes, we study the proton-emission channel of ${}^{166}$Ir $(L=2)$ and the ${}^{34}$Si-emission channel of ${}^{242}$Cm, along with their accompanying $\alpha$-decay channels. The numerical results are presented in Fig.~\ref{DCrit}. For each charged particle emitter, we consider two different assumptions on their deformations, in one of which both daughter and cluster nuclei are treated as spherical, while in the other the daughter or cluster nucleus is treated as axially deformed. We observe that the critical $D$ value beyond which the barrier penetrability gets a strong increase is given by $D_{p,\text{crit}}\approx2$ for the proton-emission channel of ${}^{166}$Ir, $D_{\alpha,\text{crit}}\approx1$ for the $\alpha$-decay channels of both $^{166}$Ir and $^{242}$Cm, and $D_{c,\text{crit}}\approx0.6$ for the ${}^{34}$Si-radioactivity channel of ${}^{242}$Cm. We also observe that, similar to $\alpha$ decay, nuclear deformations could increase the penetrability of proton emission and cluster radioactivity as well. For $\beta_{2d}=0.3$, one obtains an overall increase of around 20\% for the proton-emission channel of ${}^{166}$Ir, which is much smaller than the overall increase being around one order of magnitude for the accompanying $\alpha$-decay channel with the same $\beta_{2d}$ value. For the $^{34}$Si-radioactivity channel of ${}^{242}$Cm, we consider the deformation of the ${}^{34}$Si cluster with ${\beta}_{2c}\approx0.2$ and treat the daughter nucleus ${}^{208}$Pb as approximately spherical according to theoretical predictions in Ref.~\cite{Moller:2015fba}. We find that in this case the overall increase is more than 50 times, larger than that for the accompanying $\alpha$-decay channel.

\begin{figure}
\centering
\begin{subfigure}[b]{9.5cm}
\centering
\includegraphics[width=9.5cm]{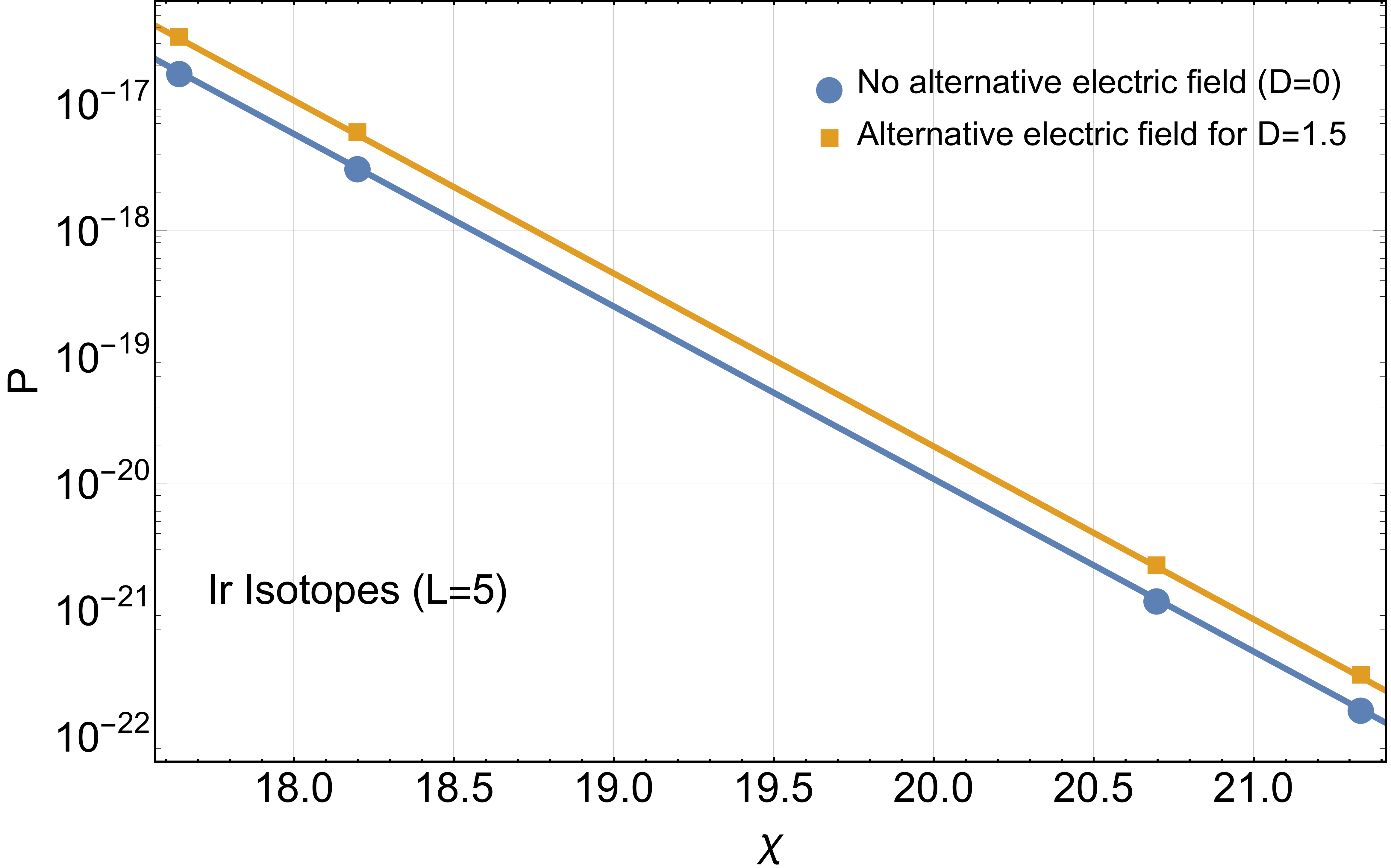}
\caption*{$\quad\quad\ \ $(a)}
\end{subfigure}

\begin{subfigure}[b]{9.5cm}
\centering
\includegraphics[width=9.5cm]{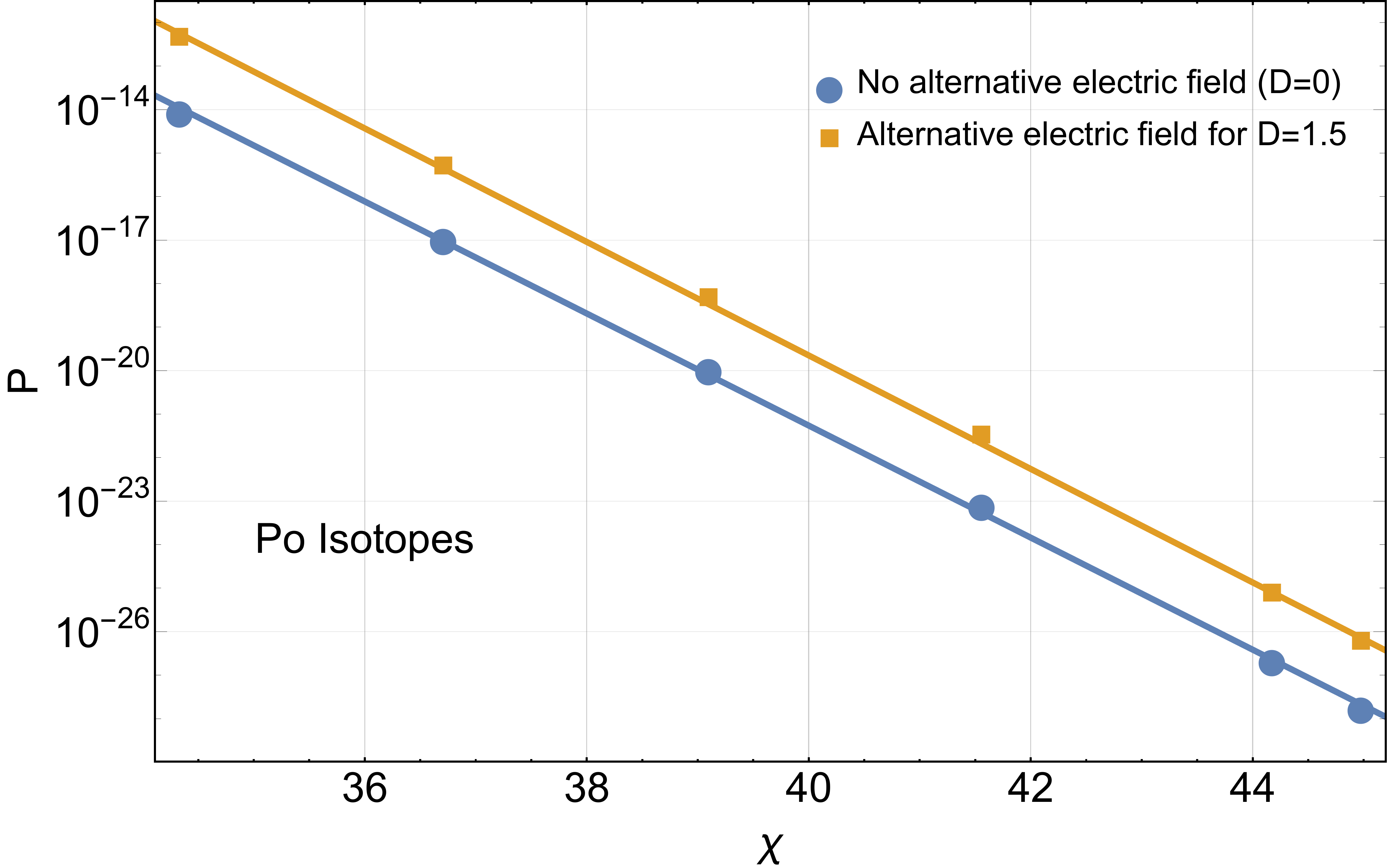}
\caption*{$\quad\quad\ \ $(b)}
\end{subfigure}

\begin{subfigure}[b]{9.5cm}
\centering
\includegraphics[width=9.5cm]{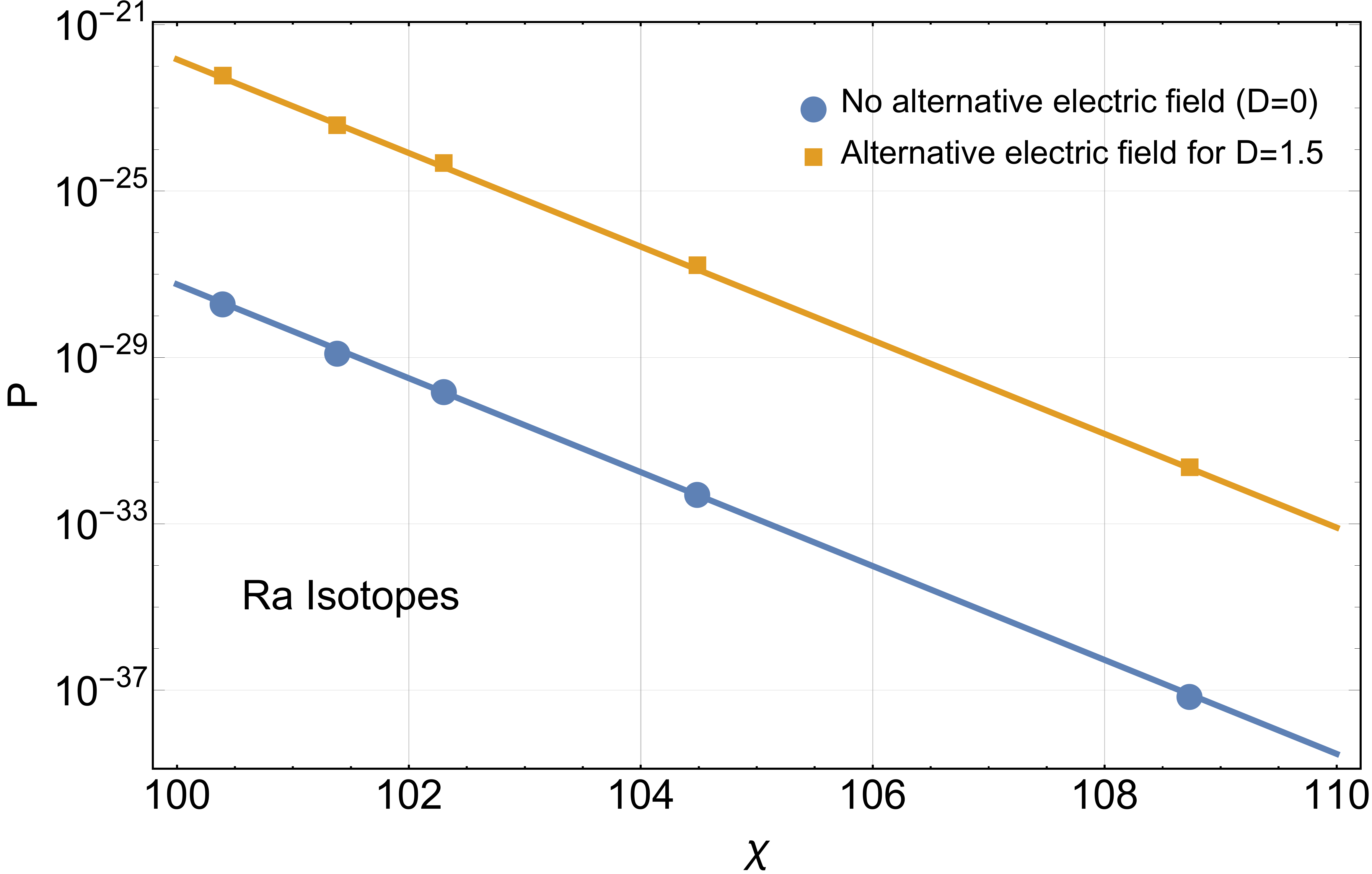}
\caption*{$\quad\quad\ \ $(c)}
\end{subfigure}

\caption{Total penetrability versus the Coulomb-Sommerfeld parameter $\chi$ for (a) Ir isotopes as proton emitters, (b) Po isotopes as $\alpha$ emitters, and (c) Ra isotope as $^{14}$C emitters in the absence of alternative electric fields $(D=0)$ and in the presence of alternative electric fields with $D=1.5$.}
\label{GeigerNuttallLaw}
\end{figure}

Furthermore, we study the shifted Geiger-Nuttall laws for proton emission and cluster radioactivity. For completeness, we reproduce here the shifted Geiger-Nuttall laws for $\alpha$ decay. We pick the ${}^{164,165,166,167}$Ir$^{(m)}$ isotopes ($L=5$), ${}^{208,210,212,214,216,218}$Po isotopes, and ${}^{221,222,223,224,226}$Ra isotopes as representative proton, $\alpha$, and cluster emitters. For simplicity, we treat all the daughter and cluster nuclei of the Ir and Po isotopes and the daughter nuclei of Ra isotopes to be spherical \cite{Moller:2015fba}, and consider the deformations of ${}^{14}$C ($\beta_{2c}=0.36$) only. The numerical results are plotted in Fig.~\ref{GeigerNuttallLaw} as the total penetrability versus the Coulomb-Sommerfeld parameter $\chi=2{Z_cZ_de^2}/{(\hbar \nu)}$, with $\nu=\sqrt{2Q/\mu}$. One could see that, similar to $\alpha$ decay, proton emission and cluster radioactivity obey the shifted Geiger-Nuttall laws as well. Given alternative electric fields of the same $D$ value, the relative corrections due to shifting terms are the smallest for proton emission, and the largest for cluster radioactivity. For instance, when $D=1.5$, Fig.~\ref{GeigerNuttallLaw} shows that the corrections induced by shifting terms are about 2 times for proton emission, 40 times for $\alpha$ decay, and five orders of magnitude for cluster radioactivity.

{
Based on the above analyses, we would like to provide some preliminary hints on the possibility that dominant decay modes of certain proton emitters could be changed by using strong high-frequency alternative electric fields. Take $^{166}$Ir as an example. In vacuum, the decay mode of $^{166}$Ir is found to be dominated by the $\alpha$-decay channel, with the branching ratios of the $\alpha$-decay and proton-emission channels given by 93\% and 7\%, respectively. When switching on strong high-frequency electric fields, according to Eq.~\eqref{QuiverAmplitude}, the $D$ values for these two decay channels are found to satisfy approximately
${D_p}/{D_\alpha}\sim15$.  
Let's take $D_\alpha\approx0.5$.
When put in such strong high-frequency alternative electric fields, as shown in Fig.~\ref{DCrit}, the total penetrability of the $\alpha$-decay channel of $^{166}$Ir remains roughly to be at the same orders of magnitude, while for the proton-emission channel we have $D_p\approx7.5$, which means the total penetrability gets an enhancement of about five orders of magnitude. Therefore, the results here indicate that, when $^{166}$Ir is placed in strong high-frequency alternative electric fields, its dominant decay mode may have the chances to be changed from $\alpha$ decay to proton emission. Similar discussions could also be extended to other proton emitters.  
}

We have also done the same analysis for cluster radioactivity, and find its branching ratio to be suppressed in general. For instance, for $^{242}$Cm, the $D$ values for the $^{34}$Si-radioactivity and $\alpha$-decay channel satisfy
${D_c}/{D_\alpha}\sim0.17$.  
Take $D_\alpha\approx3$ as an example. In this case, the total penetrability of the $\alpha$-decay channel is increased by about five orders of magnitude. Correspondingly, $D_c\approx0.5$, and the total penetrability of the $^{34}$Si-radioactivity channel is enhanced by a factor less than one order of magnitude. Therefore, in the present framework, the $^{34}$Si-radioactivity channel becomes even more suppressed in the presence of strong high-frequency alternative electric fields.

{
In summary, we study properties of $\alpha$ decay, proton emission and cluster radioactivity in the presence of high-frequency alternative electric fields, paying special attentions to the anisotropic effects, the enhancement of barrier penetrability, shifted Geiger-Nuttall laws, and the competition between different decay channels. High-frequency alternative electric fields correspond approximately to high-frequency laser fields in the dipole approximation. Ref.~\cite{Fedotov:2010,Nerush:2011,Piazza:2012,Narozhny:2015,Tamburini:2017} remark that achieving high laser intensity is not always easy in the laboratory frame. It is proposed in Ref.~\cite{Burvenich:2006zp} to use, instead, laser-nucleus collisions to achieve electromagnetic fields of high intensities directly in the nucleus rest frame. In this setup, high-frequency laser fields could be achieved naturally if the acceleration of the target nucleus is huge due to the relativistic Dopper effect. With these in mind, our study could be also viewed as a benchmark for future studies of charged particle emissions in realistic laser fields.
}

\begin{acknowledgments}
D.~B.~and Z.~R.~would like to thank Doru Delion and Stefan Ghinescu sincerely for rapid communications and valuable helps. 
This work is supported by the National Natural Science Foundation of China (Grant No.~11535004, 11761161001, 11375086, 11120101005, 11175085 and 11235001), by the National Major State Basic Research and Development of China, Grant No.~2016YFE0129300, and by the Science and Technology Development Fund of Macau under Grant No.~068/2011/A.
\end{acknowledgments}

\end{document}